\documentclass[lettersize,journal]{article}
\usepackage{amsmath,amsfonts}
\usepackage{algorithmic}
\usepackage{algorithm}
\usepackage{array}
\usepackage[caption=false,font=normalsize,labelfont=sf,textfont=sf]{subfig}
\usepackage{textcomp}
\usepackage{stfloats}
\usepackage{url}
\usepackage{verbatim}
\usepackage{graphicx}
\usepackage{cite}
\usepackage[normalem]{ulem}
\usepackage{authblk}
\usepackage[english]{babel}
\usepackage[nottoc]{tocbibind}
\usepackage{xcolor}

\begin{document}

\title{Coherent Fourier Scatterometry  for detection of killer defects on silicon carbide samples}

\author{Jila Rafighdoost, Dmytro  Kolenov, and Silvania F. Pereira}
\affil{Optics Research Group, Imaging Physics Department, Faculty of Applied Sciences, Delft University of Technology, Lorentzweg 1, 2628 CJ Delft, The Netherlands}

\maketitle





\begin{abstract}
It has been a widely growing interest in using silicon carbide (SiC) in high-power electronic devices. Yet, SiC wafers may contain killer defects that could reduce fabrication yield and make the device fall into unexpected failures. To prevent these failures from happening, it is very important to develop inspection tools that can detect, characterize and locate these defects in a non-invasive way. Current inspection techniques such as Dark Field or Bright field microscopy are effectively able to visualize most such defects; however, there are some scenarios where the inspection becomes problematic or almost impossible, such as when the defects are too small or have low contrast or if the defects lie deep into the substrate. Thus, an alternative method is needed to face these challenges. In this paper, we demonstrate the application of coherent Fourier scatterometry (CFS) as a complementary tool in addition to the conventional techniques to overcome different and problematic scenarios of killer defects inspection on SiC samples. Scanning electron microscopy (SEM) has been used to assess the same defects to validate the findings of CFS. Great consistency has been demonstrated in the comparison between the results obtained with CFS and SEM.
\end{abstract}


\section{Introduction}
Silicon carbide (SiC) has earned its place in the semiconductor industry and its applications especially in power electronic devices  \cite{1,2,3}. SiC-based devices are mainly known for their high thermal conductivity, a large bandgap, and also their high break-down electric fields  \cite{4}. SiC is transparent, but its transparency varies with wavelength. This is due to the material’s electronic structure, with the material being more transparent at longer wavelengths. At visible wavelengths, the real part of
n $\sim$ 2.6 and the imaginary part is low, indicating low absorption. However, at shorter wavelengths, the imaginary part increases. As a result, its use has been rapidly growing in power electronic devices such as aerospace materials, automotive driving, and high power converters  \cite{6,7,8}. They can be considered as the third generation of semiconductor which also happens to show a better performance compared to silicon (Si) and gallium arsenide (GaAs) as the first and second generations of semiconductors, respectively  \cite{5}.Yet, due to fabrication issues, SiC wafers might still contain a variety of defects that can put the device’s performance at risk  \cite{9,10,11,12,13,14,15}.
Defects in such semiconductor wafers can be classified into two categories - crystallographic defects found within the wafer, and morphological defects present on or near the wafer surface. The so-called killer defects that can cause device deterioration are mostly morphological and have been identified and classified  \cite{16,17,18,19,20}.

In order to avoid failures in the fabrication of devices on SiC wafers, defect inspection plays an important role. In the literature, one finds several high resolution imaging techniques that are used for inspection such as electron microscopy (TEM and SEM), atomic force microscopy (AFM), near field microscopy (SNOM). Although these techniques provide a high resolution mapping of the location and shape of the defects, there are drawbacks such as low throughput, invasiveness and they are not easily applicable for in-line inspection \cite{21,22,23,24,25,26,27}. 

These drawbacks can be avoided if one considers far field optical-based techniques such as optical coherence tomography (OCT), dark field (DF), bright Field (BF) or confocal differential interference contrast microscopy (CDIC)\cite{28,29,30,31,32}.
Among these methods, both BF and DF microscopy have proven to be the most optimal options, primarily due to their remarkable speed and cost-effectiveness
.Yet there are difficulties in applying them when it comes to  resolve the shape of very small and/or low contrast defects located on the surface or deeper within the sample. In these scenarios a complementary approach is needed to deal with the challenging defects. 

 In this paper, we present a new approach for defect characterization, namely Coherent Fourier Scatterometry (CFS) \cite{35,36,37,38}. We show that by using CFS  a complete characterization of killer defects such as stacking faults and polytype inclusions on 4H-SiC wafer can be obtained with high visibility. To validate and compare CFS with other techniques, the same sample has been inspected with BF and DF microscopy. Also, SEM was used as a calibration tool (ground truth).
 Given that this method is a far field scattering technique that uses low light power and it is not limited by diffraction, we believe that it can be considered as a complementary tool to BF and DF microscopy in scenarios like partially buried defects or isolated defects that are very small in at least one dimension such as thin lines.  In addition, since the light is focused on the sample, the scattered signal is less affected by neighboring structures as it is the case of DF microscopy.
 For defects that lie deeper into the sample,
re-focusing could be used. At last, since most of the killer
defects are of the extended type, inspection is required for
each step of the production. Therefore, the inspection needs
to be conducted throughout the process flow, meaning that an
in-line and non-destructive method is crucial. CFS also fits
these needs since it can be made very compact and easy to be
mounted in in-line tools.
 \normalsize
\section{CFS technique and experimental setup}
Scatterometry-based techniques utilize the properties of scattered light to extract information about the object being illuminated. In our implementation, we use coherent light from a laser source to illuminate the object and observe the scattered light in the far field. The details of the setup are shown in Figure 1. 
A collimated He-Ne laser with a wavelength of $\lambda$ = 632nm  passes through a non-polarising 50-50\% beam splitter (BS1). The input light is linearly polarized in  arbitrary direction before it passes through the beam splitter BS1(hereby we note that the polarization direction has negligible effect on the detection of the defects). The light then is focused on the sample plane by a high numerical aperture objective (NA=0.9). In this approach, the high numerical aperture objective allows us to illuminate the sample and collect the diffracted field in a large number of angles in one shot.

The sample is mounted on a 2D piezo-electric stage (Physik Instrumente P-629.2CD) that is programmed to be moved  in the lateral direction in raster scan mode. The focus position can also be adapted with a 1D piezo-electric stage (Physik Instrumente P-620.ZCD). Given that the focus depth is very short ($\sim 1 \mu$m), one can focus the light onto different depths within the sample.
The reflected and scattered field is captured by the same objective lens and directed back to BS1. However, due to the small depth of focus, finding the correct focal plane is challenging. We solved this issue by allowing the reflected beam from the sample to interfere with a reference beam by placing temporarily a mirror at the open port of beam splitter BS1(not shown in Figure 1). In this case, when flat interference fringes are observed between the two beams, we can ensure that the light is focused on the sample plane. The reference beam is then removed during the data acquisition. The back focal plane of the objective is imaged simultaneously into the camera and the split detector (ODD3W2 Bi-Cell Silicon Photodiode) using the telescope arrangements formed by lenses L1, L2 and L3. The camera is only used for general localization of features on the sample. The split detector is a bi-cell detector, i.e., a detector whose area is divided in two halves, with the latter being aligned perpendicular to the scan direction of one line of the raster scan. During the scan, the photocurrents of the two halves of the detector are subtracted from each other, resulting in a differential photo-current $I_{-}$ that is recorded for every scan position. After the raster scan, a 2D mapping with the values of $I_{-}$ for each scan position are plotted in a 2D graph. In order to interpret the scan maps, one can see that if there are no defects on the sample, the photo-current $I_{-}$ $\sim$ 0 (given that the scattered signal will correspond to the spurious reflection of the sample). However, when a defect passes through the focused light beam while the stage is being scanned, the far field will show asymmetries and will generate a non-zero $I_{-}$.

\begin{figure}[htbp]
    \centering
    \includegraphics[width=2.3in]{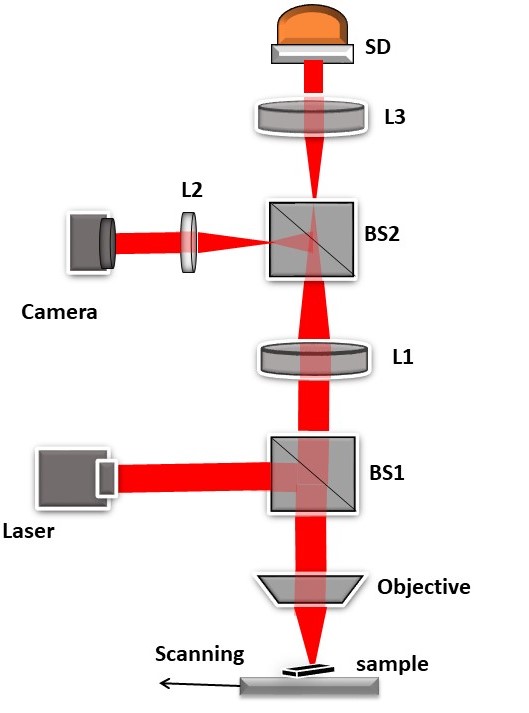}
    \caption{The schematic of the CFS setup.  The sample is placed on the stage with a piezo-electric translator, BS1 and BS2: 50-50 non-polarizing beamsplitters, L1, L2, and L3: positive lenses; Camera: CCD camera used for localizing features on the sample; SD: split detector (bi-cell silicon photodiode). A collimated and uniform He-Ne laser is used as the coherent source. The split detector is aligned perpendicular to the scan direction of one line of the raster scan.}
    \label{fig:fig1}
\end{figure}

\section{Results and Discussion}
Most defects are formed through the production process in which they not only occur on the substrate but are also generated or even modified by other layers during the wafer planarization process. 
As a result, the majority of defects can be classified as extended defects. 

Here, we measured defects on a 4H-SiC wafer with four degrees off-cut angle and 12$\pm$2 $\mu$m thickness that includes four killer defects, three stacking faults, and one polytype inclusion.
As BF and DF microscopy are the primary methodologies for general surface inspection, we have first obtained DF and BF images from a Keyence digital microscope (model VHX-6000).  Furthermore,  to verify the capabilities of CFS to detect such defects, we compared the defect shape with SEM images (Novanano, captured at 10 KV with 1000X and some with 800X magnification). In the coming subsections, we show the results. 

It should be noted that besides the observation of the aimed killer defects, some images show extra particles around the defects. This is due to the fact that the data has been captured in different labs and over different days. Unfortunately, the samples tend to accumulate dirt and contaminants from the environment. To avoid causing any damage to the samples, cleaning procedures have not been pursued. 
\subsection{Stacking Faults}
Extended defects with a planar shape are generally classified into three different groups named, stacking fault, stacking fault propagated, and stacking fault complex. Stacking faults usually appear triangular with a distant outline on the surface of SiC wafers. Propagated stacking faults have a trapezoidal structure. Our sample did not contain any stacking fault propagated and therefore we only have results for the stacking fault. Here,  Figure 2(a)-(d) show the BF and DF images, the CFS scattering map and SEM image, respectively. One can clearly see that although all methods are able to detect the defect, CFS shows a better visibility with respect to the background as compared to the other techniques. In addition, when comparing BF and DF images with CFS, the latter recovers the object even if it is not in one plane on the substrate but could extend deep into the substrate.

The next detected defect was the Stacking fault complex, which is usually called carrot. This is also a major morphological defect that has a shape  of a long and thin needle that gradually becomes even thinner at one end of the line. In Figure 3 we show images of two different carrots obtained with BF(Figure 3(a) and (c)) and DF images (Figure 3(b) and (d)).
In Figure 4 we compare the detection of the same carrot defects using CFS (Figure 4(a) and (c)) and SEM (Figure 4(b) and (d)). Similar argument with the previous defect is also valid here.
The thick head and thin tail of the defect is clearly visible with CFS and not so clear using the other techniques.

  To compare the detected signal from the defect and the surrounding background for both CFS and BF, the visibility was determined in various regions of the second Stacking Fault shown in Fig. 3(c) (BF) and Fig. 4(c) (CFS) by averaging the background ($I_{background}$) and taking the intensity of the defect ($I_{signal}$) at various points within the defect. The visibility is obtained from the expression $V$ =$(I_{defect}-I_{background})/(I_{defect}+I_{background})$. The values around the thinner tail region of the defect were $V$ = 0.06 for BF and $V$ = 0.8 for CFS. If one would take the brightest value within the defect (around the the middle of the defect), the visibility $V$ = 0.35 for BF, while the corresponding value for CFS was $V\sim$ 1. The visibilities of the other defects shown in Fig. 2, 3, and 4 are also in the same order.
\begin{figure}[htbp]
\centering\includegraphics[width=2.8in]{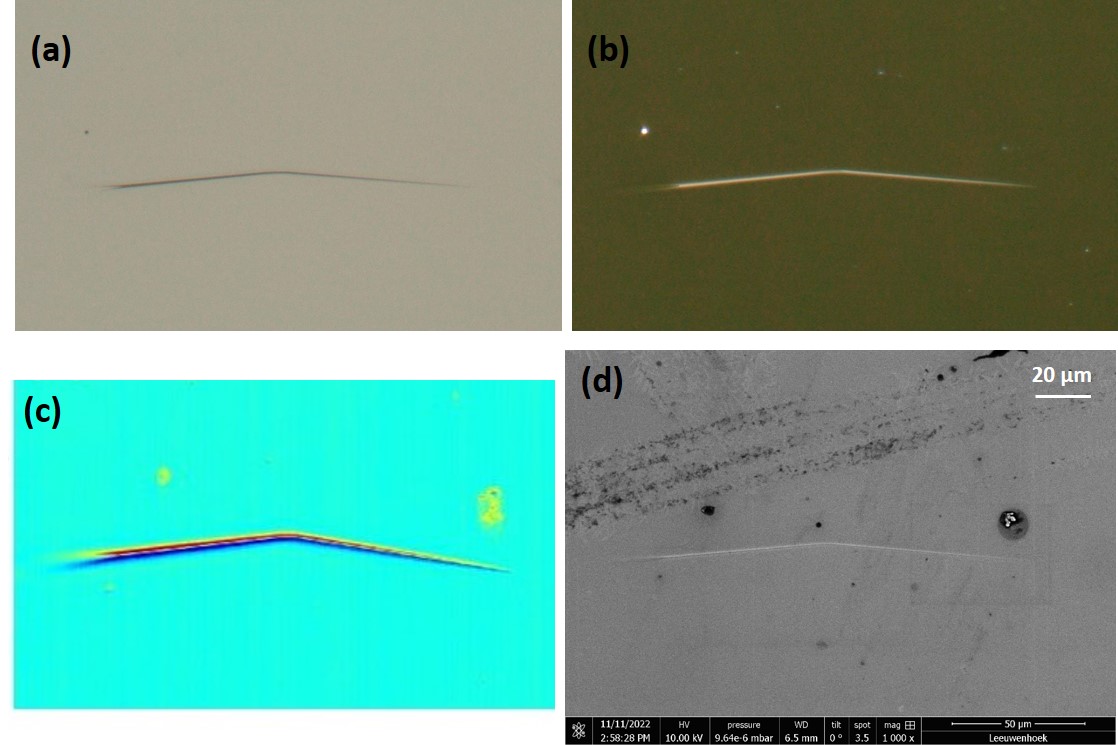}
\caption{(a) and (b) BF and DF images of the stacking fault defect, respectively. (c) CFS scan map of the stacking fault defect, and (d) SEM image of the same defect.}
\label{fig:fig2}
\end{figure}
\begin{figure}[htbp]
\centering\includegraphics[width=2.8in]{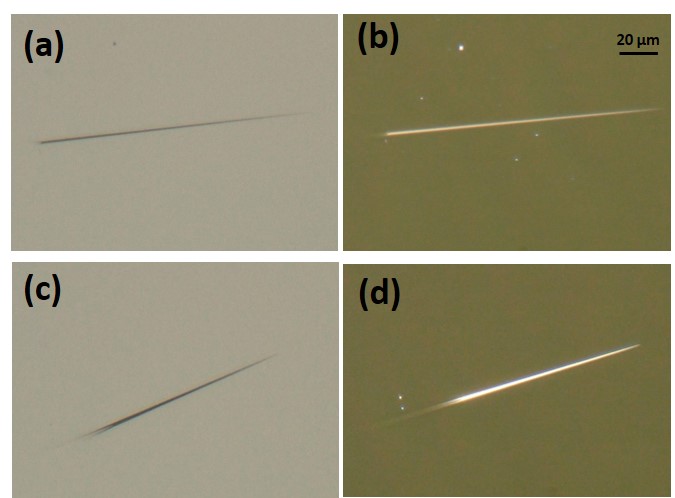}
\caption{(a) and (c) BF images of different defects, and (b) and (d) DF images corresponding to  (a)  and (c), respectively.}
\label{fig:fig3}
\end{figure}

\begin{figure}[htbp]
\centering\includegraphics[width=2.8in]{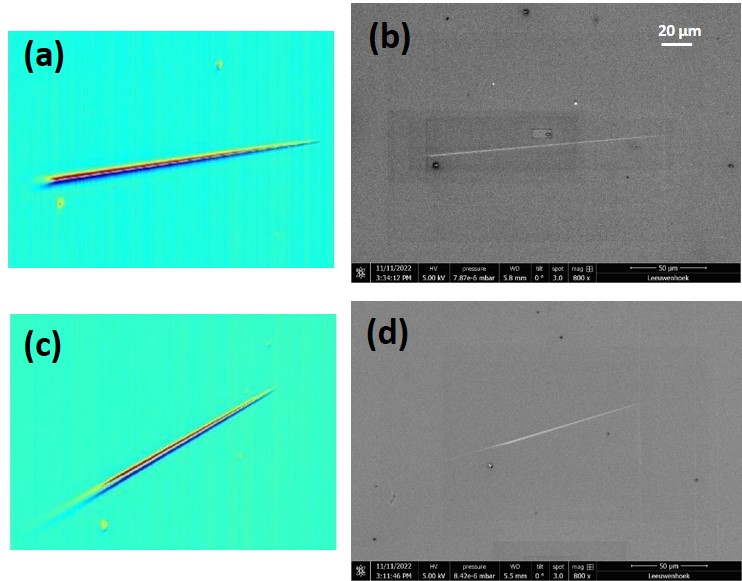}
\caption{(a)(c) and (b)(d) represent the CFS scan map (left colunm) and the SEM images (right colunm) of these defects.}
\label{fig:fig4}
\end{figure}

\begin{figure}[htbp]
\centering\includegraphics[width=2.8in]{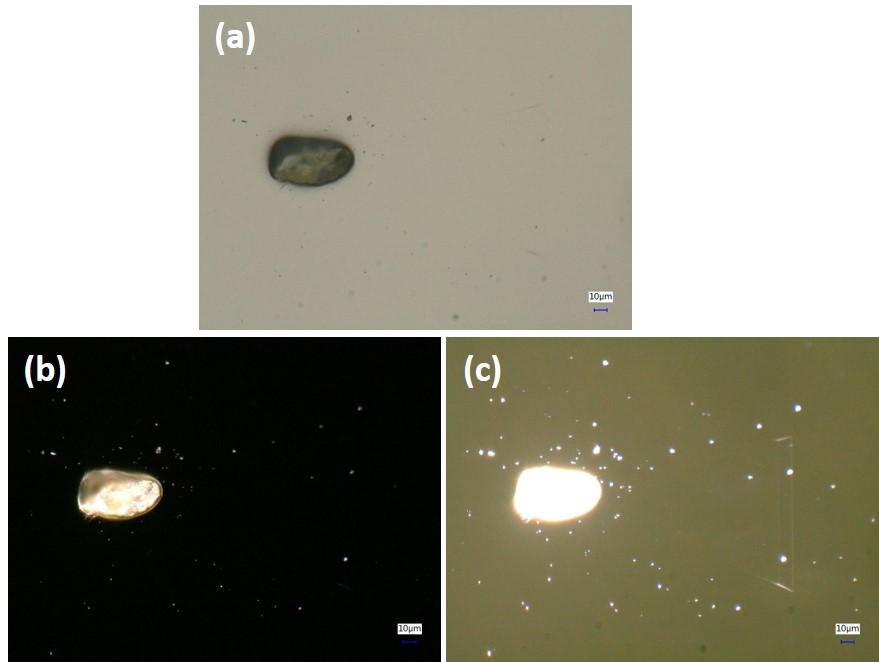}
\caption{(a) is a BF image of the defect while (b) and (c) are also DF images of the same defects. (c) is saturated in order to see the entire defect. }
\label{fig:fig5}
\end{figure}
\subsection{Polytype inclusion} 
Polytype inclusion defects are another type of morphological defect that appear as triangular with long and thin arms with a big spherical particle called downfall located at the apex of the triangle. Polytype inclusion could be formed by dropping a foreign material during the process of epitaxial growth on the SiC epilayer. 

Figure 5(a) shows the BF image of a polytype inclusion, while Figure (b) and (c) are both DF images with different powers. Figure 5(c) is saturated in order to reveal parts of the defect that are buried. Figure 6(a) shows the scattering CFS map and Figure 6(b) and (c) the SEM image of the same defect. The two SEM images are obtained at 10 and 30 KV, respectively to also highlight different parts of the defect. By comparing these images, one can see that with CFS, the triangle next to the big spherical defect is clearly revealed (see also the inset of Figure 6(a)) while  all other techniques (BF, DF and SEM) fail to image it properly. We believe that this is due to two reasons: first, the contrast was too low, and second, the triangle is in a different depth than the downfall.
This is one scenario where both optical microscopy and SEM would fail to have a comprehensive inspection due to the low penetration depth and/or low contrast. Therefore, when it comes to very narrow and features that lie partially outside the sample plane, CFS can outperform the conventional methods and become a complementary tool to them.
The comparison between the results of all techniques indeed shows good consistency and demonstrates CFS's capability in visualizing and locating killer defects.
\begin{figure}[htbp]
\centering\includegraphics[width=2.8in]{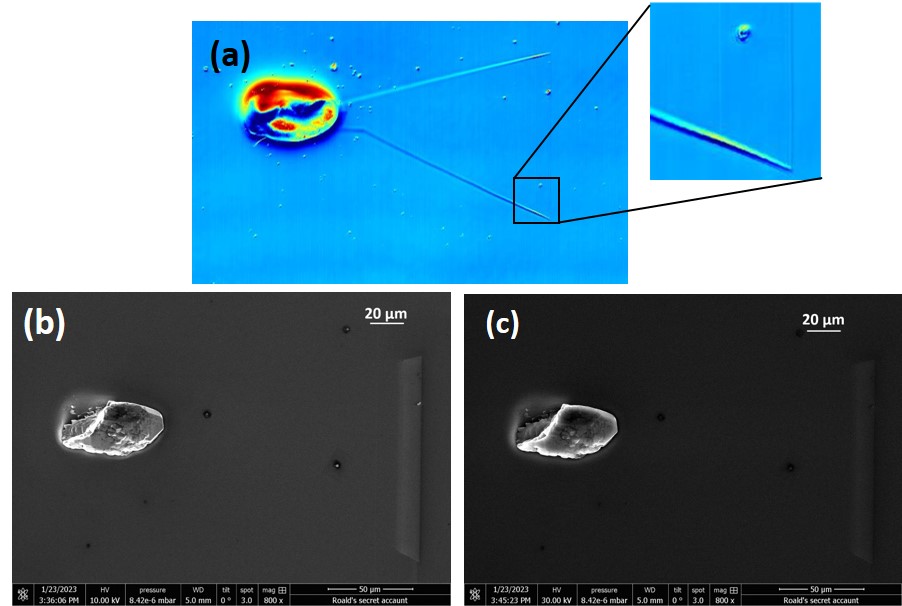}
\caption{(a) CFS scan map, (b) and (c) are SEM images of the same defect with 10 and 30 KV, respectively. In the inset in 6(a) we highlight the parts of the defect that is not clearly observed with other techniques. }
\label{fig:6}
\end{figure}
\section{Conclusion}
In this paper, we demonstrate CFS as an inspection technique for defect detection on SiC wafers that could be considered as a  complementary tool to optical microscopy. It offers some advantages including its superior visibility with respect to the background and its ability to detect defects within the substrate.

We present a proof-of-principle experiment of the technique at one wavelength (635 nm) and one numerical aperture (NA=0.9) but other wavelengths or NAs could be used as well. Shorter wavelengths/high NA are ideal for studying smaller defects, while longer wavelengths could be used to look deeper in to the sample due to longer penetration depth.
Even though CFS is not an imaging technique, we have shown that it is perfectly able to determine the shape and other details of killer defects on SiC wafers.   
The results  shown here were verified by SEM and even though the defects are visible using all techniques, CFS scattering maps provide very high visibility thanks to the differential detection scheme that highlights asymmetries of the scattered field due to the presence of the defect interacting the light beam. Furthermore, since CFS collects the scattering from different depths within the focal depth at once, it allows us to detect parts of the defect that are extended and formed in different layers in a much better way than the SEM. The major drawback so far of CFS is that it is scanning technique and it is relatively slow compared to DF or BF.
In the work we show here, the scanning was performed with piezo-electric translator and   it takes a couple of minutes to obtain an entire scan map. Thus
in order to make this technique applicable for large area inspection, further developments are needed towards fast scan combined with parallel probes techniques such as parallel optical data storage or real time confocal microscopy.
 \section*{Funding}
This work is financed by the Project 20IND09 “PowerElec” within the program EMPIR. The project is jointly supported by the European Commission and the participating countries within the European Association of National Metrology Institutes (EURAMET).

\section*{Acknowledgements}
We acknowledge AIXTRON for providing high quality state-of-the-art samples that have been used for these measurements and for fruitful discussions on different types of defects on SiC wafers. 
\bibliographystyle{unsrt}
\bibliography{biblo}


\end{document}